\documentclass[journal]{IEEEtran}

\makeatletter
\def\ps@headings{%
  \def\@oddhead{}%
  \def\@evenhead{}%
}
\makeatother

\IEEEoverridecommandlockouts
\makeatletter
\def\ps@IEEEtitlepagestyle{%
    \def\@oddhead{}%
    \def\@evenhead{}%
}
\makeatother

\usepackage{comment}

\ifCLASSINFOpdf
  \usepackage[pdftex]{graphicx}
\else
   \usepackage[dvips]{graphicx}
   
\fi

\begin{document}
%
\title{\textsc{Drama Llama}: An LLM-Powered Storylets Framework for Authorable Responsiveness in Interactive Narrative}
%
%
%

\author{Yuqian~Sun,
        Phoebe~J.~Wang,
        John~Joon~Young~Chung,
        Melissa~Roemmele,
        Taewook~Kim,
        and~Max~Kreminski
\thanks{Y. Sun, J. J. Y. Chung, M. Roemmele, and M. Kreminski are with the Midjourney Storytelling Lab, e-mails: ysun@midjourney.com, jchung@midjourney.com, mroemmele@midjourney.com, mkreminski@midjourney.com}
\thanks{P. J. Wang is with sadstory.gg, e-mail: wjingsen@gmail.com}
\thanks{T. Kim is with Northwestern University, e-mail: taewook@u.northwestern.edu Work conducted while at Midjourney.}
\thanks{Manuscript received December 16, 2024.}}

%
%

\markboth{IEEE Transactions on Games,~Vol.~XX, No.~X, May~2025}%
{Shell \MakeLowercase{\textit{et al.}}: Bare Demo of IEEEtran.cls for IEEE Journals}
%



\maketitle

\begin{abstract}
In this paper, we present Drama Llama, an LLM-powered storylets framework that supports the authoring of responsive, open-ended interactive stories. DL combines the structural benefits of storylet-based systems with the generative capabilities of large language models, enabling authors to create responsive interactive narratives while maintaining narrative control. Rather than crafting complex logical preconditions in a general-purpose or domain-specific programming language, authors define triggers in natural language that fire at appropriate moments in the story. Through a preliminary authoring study with six content authors, we present initial evidence that DL can generate coherent and meaningful narratives with believable character interactions. This work suggests directions for hybrid approaches that enhance authorial control while supporting emergent narrative generation through LLMs.
\end{abstract}


\IEEEpeerreviewmaketitle

\section{Introduction}

\IEEEPARstart{A}{uthors} of interactive stories are subject to two major forces, which pull in such strongly opposed directions that the navigation of the resulting tension forms the foundation for a whole field of narrative systems research. The first of these forces is the desire of authors to provide the player with \emph{responsiveness}: the impression of a narrative experience that is ``intelligently responding to a player and adapting to them''~\cite{Responsiveness}. The second is the desire to limit \emph{authorial burden}~\cite{AuthorialBurden}: the growing difficulty of writing and managing a sufficient corpus of proceduralized narrative content as the range of potential player trajectories to which the system must respond expands.

One increasingly popular way to structure interactive story content is through the translation of potential story events into \emph{storylets}~\cite{Storylets}: discrete and self-contained units of narrative, each one consisting of \emph{preconditions} (which permit the storylet to be presented to the player under certain game state conditions); \emph{content} (which is made available to the player when the storylet is selected for presentation); and potentially \emph{effects} (which update the game state in response to the storylet's selection or the player's traversal of its content). Storylet structures allow for the gradual expansion of a narrative system's responsiveness through the addition of storylets that match progressively more specific combinations of game state conditions; the modularity of storylets may also make storylet-based interactive narratives easier to narratively ``refactor'' than their linear or branching counterparts.

Nevertheless, storylet systems bring authoring burdens of their own. Most notably, the use of sophisticated logical preconditions to match complex combinations of game state conditions can make storylets rather finicky to author~\cite{ExperiencingAuthorialBurden}, especially for less-experienced procedural writers: defining, encoding, debugging, and refining the exact logic governing a storylet's availability can take up the bulk of authoring time for creators of modularized narrative content. Moreover, writing a storylet-based interactive narrative requires the up-front enumeration of a closed ontology of narrative event types that can occur---potentially limiting the player's capacity to take the story in directions that the story's author did not originally consider, even if those directions are consistent with the author's extrapolated vision~\cite{Spleenwort}.

Purely LLM-based narrative systems can be viewed as addressing these problems: narrative content is authored as natural language (e.g., introductory setting descriptions and descriptions of character personalities), and the system is free to improvise responses to arbitrary natural language player input. However, LLM-based systems also tend to struggle with narrative structurelessness (narrative events occur without driving the story toward any particular progression) and lack of pushback (it's hard to get the system not to go along with whatever the player suggests, even if those ideas are totally out of line with the author's vision).

Can the strengths of storylet- and LLM-based approaches to interactive storytelling be combined? In this paper, we present \textsc{Drama Llama}: an LLM-powered storylets framework that brings together the high-level content structure of storylet-based systems with the natural language authoring of purely LLM-based systems. Our aim is to support the creation of responsive, open-ended interactive stories that can recognize and elaborate on player-introduced narrative ideas while driving playthroughs toward author-crafted narrative goals.

\section{Related Work}

\begin{table*}[]
\renewcommand{\arraystretch}{1.3}
\begin{tabular}{lllllll}
                            & \begin{tabular}[c]{@{}l@{}}Authoring \\ Tool\end{tabular} & \begin{tabular}[c]{@{}l@{}}Hierarchical\\ decomposition\end{tabular}      & \begin{tabular}[c]{@{}l@{}}Neural/LLM \\ generation\end{tabular} & Allow storylets & Emergent narrative                                                                       & \begin{tabular}[c]{@{}l@{}}Open-ended \\ generation\end{tabular} \\ \hline
\emph{Façade}~\cite{mateas2003faccade}                           &                                                           & Y                                                                         &                                                                  &                 &                                                                                          &                                                                            \\
Versu~\cite{Versu}                            & Y                                                         & Y                                                                         &                                                                  & Y               &                                                                                          &                                                                            \\
StoryNexus~\cite{storynexus}                       & Y                                                         & Y                                                                         &                                                                  & Y               & Y                                                                                        &                                                                            \\
StoryAssembler~\cite{StoryAssembler}             & Y                                                         & Y                                                                         &                                                                  & Y               &                                                                                          &                                                                            \\
Spleenwort~\cite{Spleenwort}                         & Y                                                         & Y                                                                         & Y                                                                &                 &                                                                                          &                                                                            \\
StoryVerse~\cite{storyverse}                  & Y                                                         & Y                                                                         & Y                                                                &                 & Y                                                                                        &                                                                            \\
Generative Agents~\cite{GenerativeAgents}            &                                                           &                                                                           & Y                                                                &                 & Y                                                                                        &                                                                            \\
c.ai~\cite{characterai} & Maybe& Maybe, with complex prompts& Y                                                                &                 &                                                                                          & Y                                                                          \\
ChatGPT~\cite{ChatGPT}                     & Maybe& Maybe, with complex prompts& Y                                                                &                 &                                                                                          & Y                                                                          \\
AI Dungeon~\cite{AIDungeon}                  & Maybe& Maybe, with complex prompts& Y                                                                & Maybe& Maybe, with enough contents& Y                                                                          \\
Dramatron~\cite{dramatron}                   & Y                                                         & Y                                                                         & Y                                                                &                 &                                                                                          &                                                                            \\
Drama Llama                 & Y                                                         & Y                                                                         & Y                                                                & Y               & Maybe, with  proper storylets& Y                                                                         
\end{tabular}
\caption{Comparison of Selected Generative Narrative Systems}
\label{table:compare}
\end{table*}

Storylets are increasingly widely used to structure content for both open-ended emergent narrative experiences~\cite{IENHistory} and drama-managed experiences~\cite{DramaManagement,EvaluatingDramaManagement} in which a computational agent actively pursues the construction of interactive stories. Existing storylet systems---including StoryNexus~\cite{storynexus}, Versu~\cite{Versu}, StoryAssembler~\cite{StoryAssembler}, and Praxish~\cite{Praxish}---rely on authors to specify fine-grained units of narrative content that are unlocked based on certain conditions. For instance, a character in a mystery story may need to reach an ``angry" state before they can trigger the storylet that accuses the murderer. This approach can enable high responsiveness to player input while preserving authorial intent. However, as Emily Short notes, storylet-based authoring faces significant scaling challenges: ``The content tends to be uninteresting until there are a fair number of storylets in the database [...] it's hard to feel like you're really rolling until you've spent quite a bit of time in the tool"~\cite{short2016beyond}.

A library of storylets constitutes a closed ontology of narrative actions that may occur, which can limit the responsiveness of storylet-based interactive stories to unexpected player input. Some efforts have been made to mitigate this, for instance by using various NLP techniques to support player discovery of actions that match their natural-language expressions of intent~\cite{mateas2003faccade,TextToDialog,PWIM,Avrae}. However, these input-level approaches do not fundamentally change the ``closed-world'' nature of explicitly enumerated storylet libraries.  

On the other hand, LLM-based systems can generate narrative content with less up-front authorial specification, but preserving a coherent authorial intent remains challenging. Recent LLM-based approaches to narrative generation fall into two broad categories, roughly analogous to the ``bottom-up'' (emergent narrative) and ``top-down'' (planning-based) approaches taken by symbolic AI before them~\cite{BalancingPlotAndCharacter,NarrativePlanning}. Emergent narrative systems, such as AI Dungeon~\cite{AIDungeon}, generate interactive narratives from minimal world and character specifications;  in such systems, character dialogues and behavior descriptions tend to be generated entirely at runtime. While this approach reduces the authoring burden, it can also result in content homogeneity~\cite{homogeneity} and limited authorial control. Similarly, agent-based frameworks (e.g., Park et al.'s ``generative agents''~\cite{GenerativeAgents}) showcase the unpredictable emergent behaviors of simulated characters but do not provide strong authorial affordances for shaping character behavior. 

The alternative approach, hierarchical decomposition---seen in systems like StoryVerse~\cite{storyverse}, Spleenwort~\cite{Spleenwort} and Dramatron~\cite{dramatron}---attempts to guide LLM generation of story text through high-level author-specified outlines of story structure. This provides a means of high-level structural control, but not necessarily of fine-grained event management. Additionally, the rigid high-level structures to which hierarchical systems adhere may make it difficult to adapt them to player input; systems in this category are more often deployed as end-to-end story generators than as responsive interactive stories.

Yang et al.~\cite{GPTForGames} provide an overview of many recent LLM-based narrative play experiences, including storytelling games like \emph{1001 Nights}~\cite{1001Nights}; community storytelling chatbots~\cite{FictionalWorldsRealConnections}; and LLM-powered ``dungeon masters"~\cite{DnDChallenge}. One common challenge faced by all of these systems is how to balance player responsiveness (e.g., they should be able to talk to characters) with authorial intention (e.g., characters should not say out-of-context sentences)~\cite{kim2024authors}.

Altogether, these related works inspired us to mix storylet systems with LLM-driven generation to support a responsive and interactive hybrid authorial system. Table~\ref{table:compare} summarizes key similarities and differences between some relevant prior generative narrative systems.

\begin{figure*}
    \centering
    \includegraphics[width=0.8\linewidth]{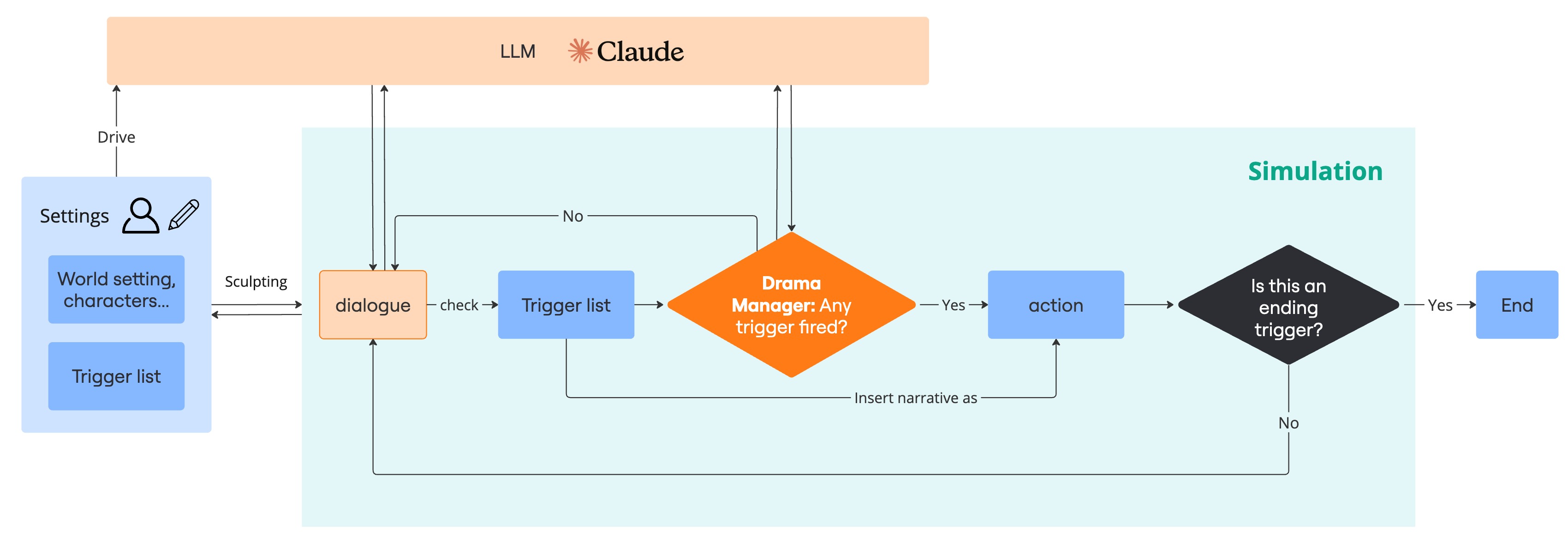}
    \caption{Overall diagram of the \textsc{Drama Llama} system.}
    \label{fig:enter-label}
\end{figure*}
\section{System Description}

\textsc{Drama Llama} is an authoring framework that leverages storylets and LLMs to support the creation of responsive, open-ended text-based interactive stories. The authoring process is iterative: authors define a story setting, refine it based on observations of simulated character behaviors, then progressively add and adjust storylets based on emerging outputs.

\begin{figure*}
    \centering
    \includegraphics[width=0.85\linewidth]{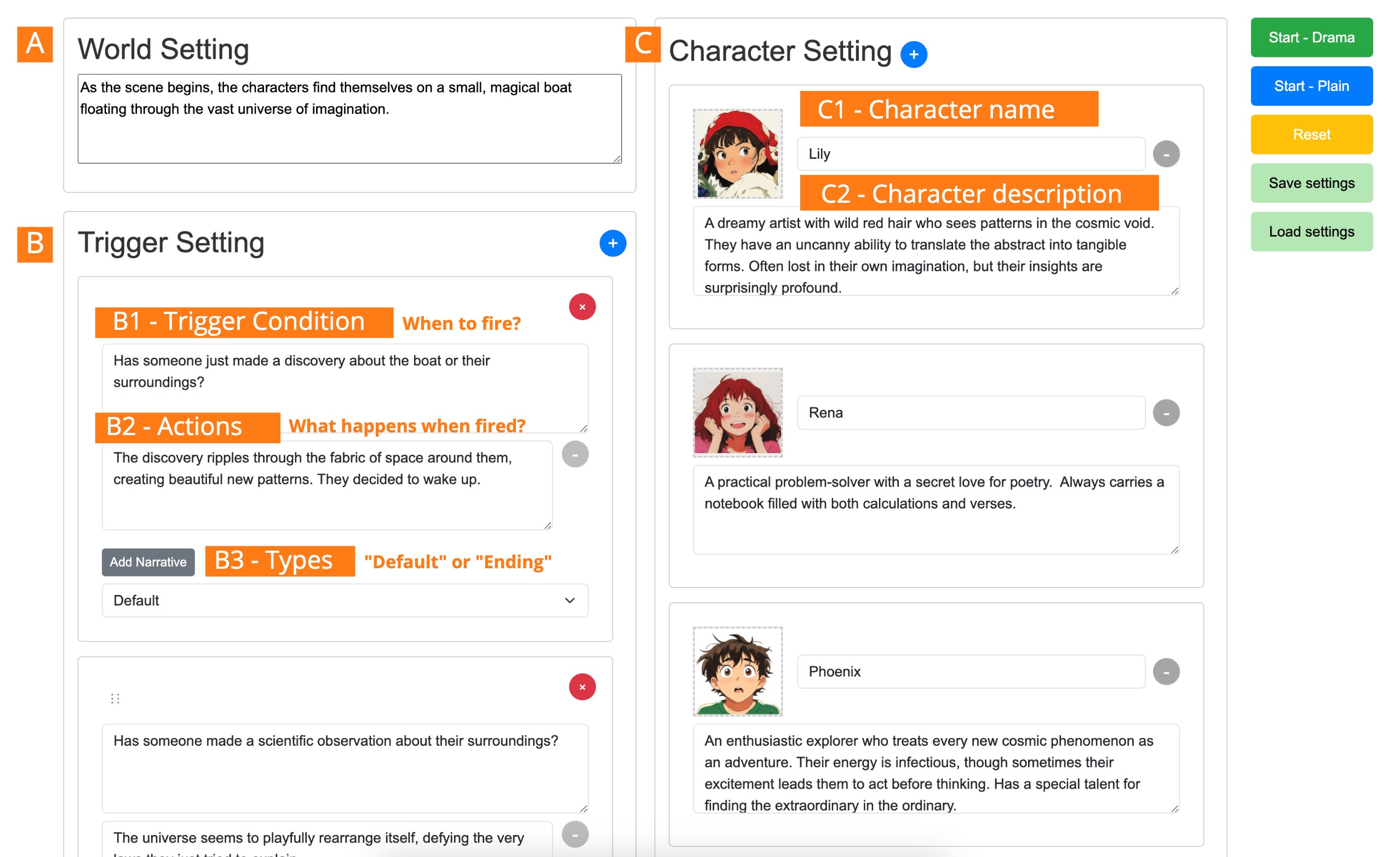}
    \caption{The setting configuration interface used by authors to define their story setting, characters, and triggers.}
    \label{fig:setting}
\end{figure*}

Broadly speaking, a \textsc{Drama Llama} story is defined in terms of a few key human-authored elements:
\begin{itemize}
\item A natural-language \textbf{world setting description} that initializes the broad setting of the story. (Fig.~\ref{fig:setting}A) 
\item A cast of \textbf{characters}. Each character has a name; a short player-visible description; and a natural-language prompt that describes their personality and behavior. (Fig.~\ref{fig:setting}C)
\item A set of \textbf{triggers}, analogous to storylets. An example trigger can be found in Appendix~\ref{sec:exampletriggers}.
\end{itemize}

Each trigger in turn consists of:
\begin{itemize}
    \item A \textbf{condition}: A natural-language definition of the condition under which the trigger should fire (Fig.~\ref{fig:setting}B1).
    \item \textbf{Action(s)}: A sequence of natural-language pieces of ``stage direction'' text to inject into the player-visible story when the trigger fires (Fig.~\ref{fig:setting}B2). Each time the trigger condition is met, the next unused action in the sequence is injected into the story and becomes inactive. 
    The trigger itself becomes inactive once all actions are consumed.
    \item A trigger \textbf{type}: either ``Basic" or ``Ending" (Fig.~\ref{fig:setting}B3). When an Ending trigger is fired, the simulation  stops.
\end{itemize}

During gameplay, characters are portrayed by LLM agents, reacting to what other characters have said and done. One character is controlled by the player; when the system generates dialogue for the player-controlled character, this dialogue is discarded and the player is instead prompted to type in what their character says next.

\begin{figure*}
    \centering
    \includegraphics[width=0.85\linewidth]{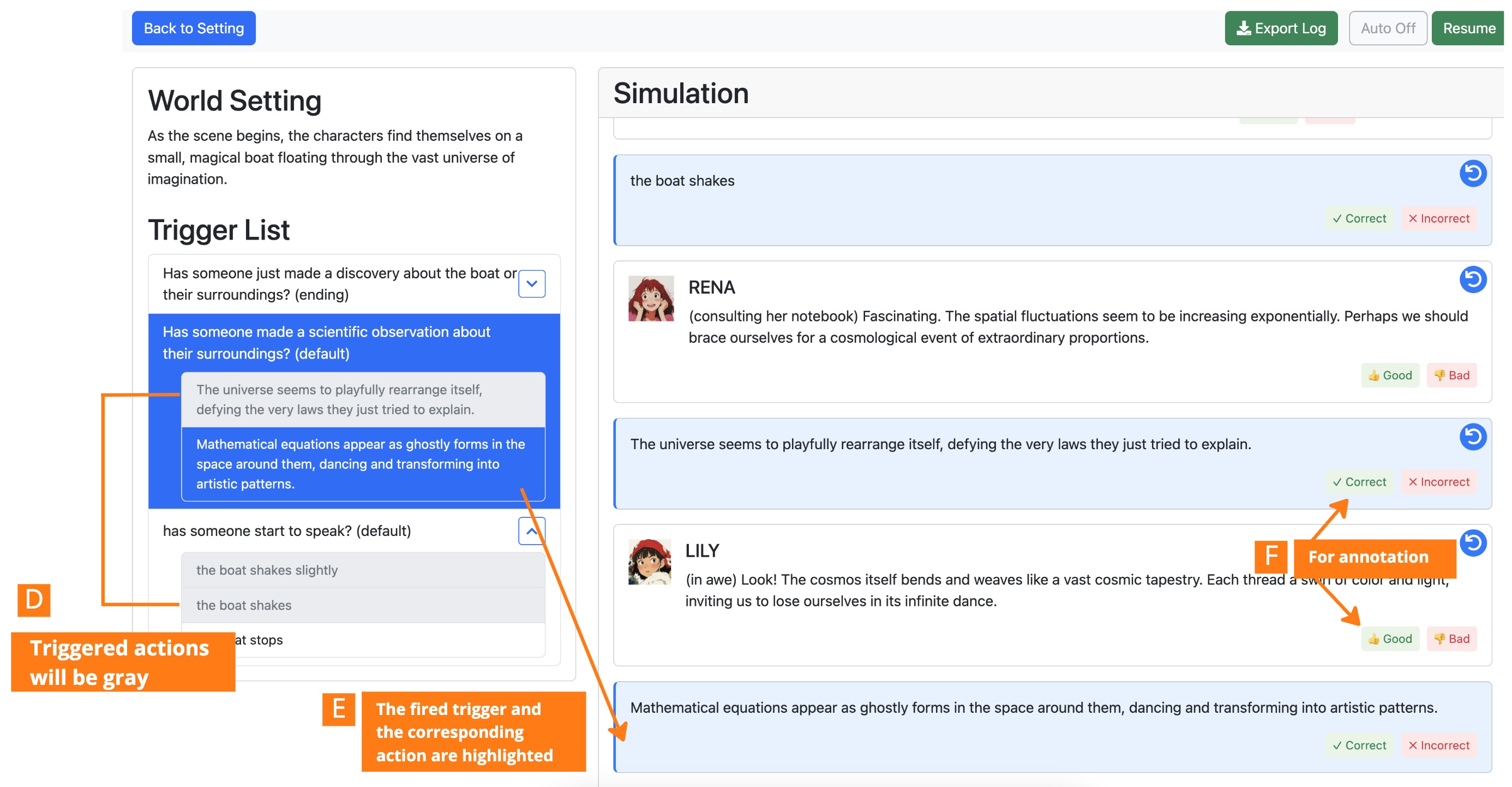}
    \caption{The simulation interface used by authors to test their story definition as they write and modify the setting and trigger content.}
    \label{fig:simulation}
\end{figure*}

After every message, a minimal LLM-based \emph{drama manager} decides which trigger (if any) to fire by sequentially checking the conditions of each active trigger against the story text so far, stopping when it finds a trigger to fire. When a trigger fires, its next unused action text (Fig.~\ref{fig:simulation}E) is appended to the story, so that the player and simulated characters may react to it. System prompts are attached as Appendices~\ref{sec:simprompt} and \ref{sec:triggerprompt}.


The work of authoring a \textsc{Drama Llama} story, then, centers predominantly on identifying potential \emph{transitions} that might take place within the storyworld (for instance between ``scenes'', ``acts'', or other distinct sections of a larger narrative arc) and defining trigger rules that will fire to enact these transitions at appropriate times. In order, authors:

\begin{enumerate}
\item Define initial settings (Fig.~\ref{fig:setting})
\item Run the characters in autonomous mode for a few turns (Fig.~\ref{fig:simulation}), over and over again, to get a feel for what they tend to do unassisted
\item Sculpt initial settings based on observations, so the story tends to go in ``good'' directions per the author's taste
\item Once satisfied with \emph{default} character behavior paths, define triggers that fire when the story progresses in certain typical directions
\item Repeat process to sculpt triggers, follow-up triggers, polish character settings, etc.
\end{enumerate}

\section{Evaluation}
\begin{figure*}
    \centering    \includegraphics[width=0.7\linewidth]{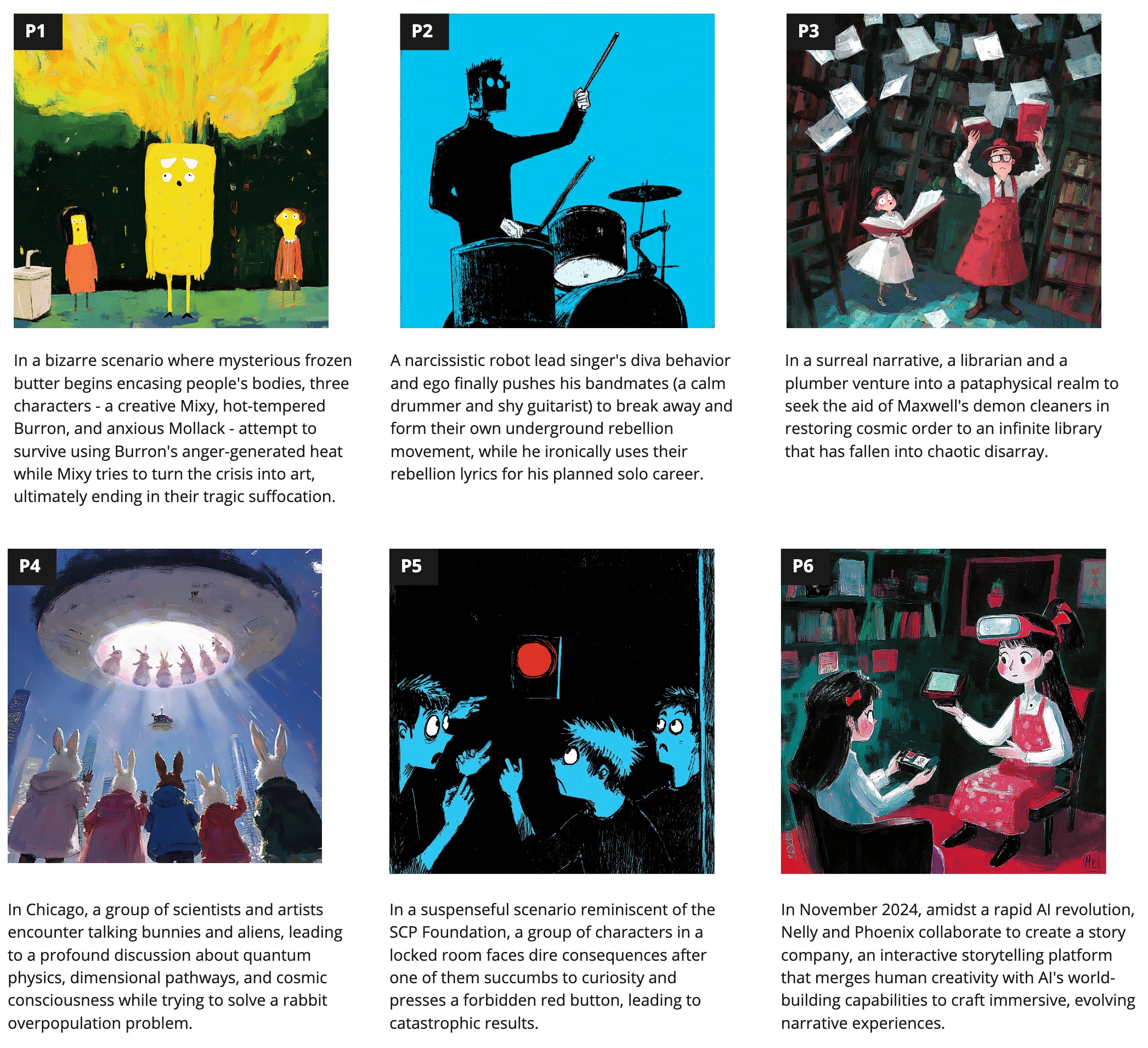}
    \caption{Textual and visual summaries of the stories produced for Task A by our six authoring study participants.}
    \label{fig:stories}
\end{figure*}

As a preliminary evaluation of \textsc{Drama Llama} (DL), we conducted a small authoring study involving six participants with extensive experience in interactive storytelling. Three participants were men and three were women; their ages ranged from 27-34. The study was conducted remotely, with participants receiving detailed guidelines, access to the DL web interface, and evaluation questionnaires. Participants were given flexibility to complete the tasks at their convenience. Throughout the study, they could communicate with the researchers via Discord for clarification and technical support.

\subsection{Task A: Open-Ended Authoring}
Participants were first instructed to create an open-ended interactive story using DL. Participants had complete creative freedom in establishing the world settings, characters, and narrative triggers. They could run and pause the simulation, reset to previous states, or modify settings at any point during the creation process. Upon achieving a satisfactory playthrough of their story, participants were asked to:
\begin{enumerate}
    \item Annotate triggers and character dialogues (Fig. \ref{fig:simulation}F). Participants were required to evaluate all trigger activations for \textbf{Trigger Accuracy}, indicating whether triggers fired at appropriate times within the narrative context. Participants also optionally annotated generated lines of dialogue for \textbf{Dialogue Quality} to identify particularly effective (``good'') or problematic (``bad'') lines. 
    \item Self-evaluate the annotated narrative via the Torrance Test for Creative Writing (TTCW) questionnaire. TTCW~\cite{ttcw} is a robust test for assessing creativity in fictional short stories. The test consists of 14 binary tests organized into the original dimensions of Fluency, Flexibility, Originality, and Elaboration. 
    \item Provide open-ended writing feedback on the narrative.
\end{enumerate}

\subsection{Task B: Fixed-Topic Authoring}

To enable direct comparison between different participants' usage of DL, participants were then tasked with authoring an interactive story based on a predetermined scenario. To incorporate inherent dramatic tension, we drew inspiration from \emph{Façade}~\cite{mateas2003faccade}, a seminal work in narrative intelligence and interactive drama. The scenario---``As an invited old friend, the protagonist intervened in a couple's argument"---was adapted from \emph{Façade}'s core dramatic premise and provided to participants as a fixed topic. Following the structural requirements, participants were instructed to include a minimum of three characters: two characters forming the couple and one character as the intervening friend. Similarly to Task A, participants first constructed their story definition, then produced and self-evaluated a single output transcript.

Throughout all tasks, participants interacted with the system through a web-based interface. Data collection included exported narratives, annotations of significant story elements, and structured feedback through questionnaires. 

\begin{table*}[]
\centering
\begin{tabular}{llll|lll}
   & Characters  & Triggers    & Action per trigger & Simulation length & Dialogues     & Actions     \\ \hline
A  & 3.50 ± 1.26 & 3.33 ± 2.36 & 1.53 ± 0.76        & 28.33 ± 13.14           & 22.00 ± 12.48 & 6.33 ± 3.54 \\
B & 3.33 ± 0.47 & 4.00 ± 2.52 & 1.38 ± 0.69        & 25.33 ± 16.38           & 20.67 ± 16.50 & 4.67 ± 1.89 \\
\end{tabular}
\caption{Summary of Authored Stories and Representative Playthroughs}
\label{table:statistics}
\end{table*}

\section{Results}

\subsection{Torrance Tests for Creative Writing (TTCW)}


Among properties of creative writing evaluated by the TTCW (Appendix \ref{ttcw result}), authors most frequently agreed that DL outputs showed \textit{structural flexibility} (5 authors in Task A and all 6 in Task B), indicating DL's ability to maintain coherent narrative structures. DL also performed well at \textit{narrative pacing} (4 in Task A, 5 in Task B), \textit{understandability \& coherence} (3 in Task A, 6 in Task B), \textit{emotional flexibility} (6 in Task A, 3 in Task B), and \textit{world building \& setting} (4 in Task A, 5 in Task B).

However, for both tasks, only 3 authors agreed that outputs showed \textit{originality in thought}. Two authors specifically noted that their stories felt ``clichéd," with one attributing this to insufficient character detail in the initial setup. In terms of \textit{character development}, scores were also low (2 in Task A, 3 in Task B), possibly due to what P4 author described as "overly forced” alignment of outputs with initial character settings: system outputs often repeatedly re-emphasized characters' prescribed traits.

These findings suggest that while DL is helpful in maintaining narrative structure and coherence, it is currently still limited in generating original thoughts and more nuanced character development.

\subsection{Author Feedback}

Analysis of author feedback reveals both strengths and weaknesses of the current version of DL, as well as potential directions for future research and development.

\textbf{DL demonstrated capability in generating coherent and meaningful narratives with believable character interactions.} P2 noted that ``the resulting story was actually very meaningful and had a clear plot," while P3 expressed surprise and please at "the coherence" of the generated content. P5 specifically highlighted the system's ability to maintain believable character reactions and manage narrative tension.

However, three of six participants identified a tendency toward clichéd content. As P3 observed, ``it was quite clichéd," though P2 qualified this criticism by noting that ``the story does have many cliches, but that isn't necessarily bad." System stochasticity emerged as another concern, particularly regarding trigger timing. P1 reported that "detection of triggers did not work as expected". 

The cognitive load of maintaining trigger consistency emerged as a significant consideration. P1 mentioned: ``I too much focused on 'trigger consistency'... I was kind of spending too much of my cognitive efforts to them, instead of other qualities of the tool."

\textbf{Feedback revealed that the quality of generated drama largely depended on the efforts of author input.} P3 observed that DL was ``good at following directions," noting stronger performance in tasks with detailed prompts. This observation was reinforced by P6's self-reflection: ``I want it to be more dramatic, but I was too lazy to write the triggers. I guess the responsibility is on me." Similarly, P5 mentioned that although she liked the tension between characters, she hoped the conversations (and tension) could be longer. She said, ``Perhaps due to my direct explanation of triggers, the scenes were fairly short."

These findings align with our TTCW results, suggesting that while DL's constraint to pre-written triggers may limit spontaneity, it provides authors with greater narrative control when properly utilized.

\subsection{Comparison of Authored Stories}

An initial analysis of the narratives produced for Tasks A and B reveals different patterns in authorial approaches and thematic emphases (Fig.~\ref{fig:stories}). In Task A, the authors demonstrated a propensity toward heightened dramatic tension and fantastical elements, like the robot band's interpersonal dynamics (P2), the librarian's metaphysical journey (P3), and the surreal rabbit invasion narrative (P4). Notably, all works incorporated sci-fi and surreal backgrounds, which drove the system to produce surprising and inspiring content. As P2 noted, the system coherently generated genre-appropriate humor, such as robots calling each other ``pile of scrap" and suggesting to ``reboot attitude subroutines."

In contrast, Task B narratives demonstrated a focus on nuanced interpersonal and emotional mediation within realistic domestic contexts. Only P4 maintained a sci-fi background similar to his Task A setting, adapting it to the visitor-arguing couple relationship format. Notably, across Task B stories, the mediator characters tended to employ increasingly rational argumentation and logical persuasion as stories progressed. They shifted towards reasoned discussion rather than overheated arguments. P4 critiqued this aspect of character emotion, noting: ``Maybe the emotion aspect is way too much flexible? Sometimes the flexibility seems even unrealistic. After a huge argument or fight, people would not be resilient that good."

\section{Discussion}

DL suggests a balanced approach between authorial control and emergence. The system inherits storylet-driven thinking while leveraging LLM capabilities to reduce the implementation burden. Authors need only define a small number of triggers to drive dramatic interaction. Natural language descriptions replace the complex precondition logic of traditional storylet systems. This balance allows authors to exert precise event-level control where they want it while enabling emergent content within the prescribed framework.

This approach suggests a shift in the authoring mindset. Rather than explicitly enumerating all narrative possibilities alongside rules that govern their assembly (as with traditional storylet systems) or specifying complete story trajectories (as in hierarchical decomposition), authors of DL stories instead define key narrative possibilities that can serve as ``pivot points'' for stories made of mostly improvised textual material. Our study demonstrates that DL writers can focus on characters and events of interest, with an average of 3-4 well-written triggers (as shown in Table~\ref{table:statistics}) generating engaging drama simulations. The system manages detail generation within the prescribed framework, balancing control and autonomy.

In interactive storytelling theory, DL explores two dimensions identified by Riedl and Bulitko~\cite{riedl2013interactive}: authorial intent and virtual character autonomy. The storylet structure maintains author control over key events, while LLM implementation enables character autonomy within the prescribed framework. This balance positions DL as a computational caricature~\cite{caricature}, demonstrating one approach to maintaining authorial intent while leveraging AI generative capabilities in the LLM era. A more full-fledged version of the system could eventually support RPG or visual novel experiences where player responses remain well-aligned with authorial intentions.

It should be noted that natural-language authoring may not be a panacea for content authorability: the authoring of effective natural language prompts can be difficult for inexperienced users \cite{WhyJohnnyCantPrompt}, and the same prompt may yield very different results when used with different LLMs~\cite{StateOfWhatArt}. However, a study of end-user authoring of \emph{classifier rules} that discriminate positive from negative examples has also found natural language to be advantageous over rule authoring in a more procedural language for rapid convergence on higher-performance classifiers \cite{EndUserClassifierAuthoring}. In conjunction with our observation that authors generally picked up the authoring of DL stories quite rapidly, we believe this suggests strong potential authorability benefits for natural-language triggers over purely procedural alternatives.

\section{Limitations and Future Work}
The current implementation of \textsc{Drama Llama} prioritizes system responsiveness by processing triggers sequentially (and discontinuing trigger list processing after a single trigger is fired) rather than evaluating all options at each step. This design choice enables real-time generation within 3 seconds per step and allows authors to arrange trigger order in their settings. While this approach achieves satisfactory performance with minimal computational overhead, several improvements remain for future development.

We plan to enhance the system's capabilities in two main directions. First, we plan to expand the range of trigger types to support more diverse narrative scenarios, such as:
\begin{itemize}
\item \emph{Fallback} triggers that fire whenever no other triggers have fired for at least $K$ successive lines, allowing authors to impose a ``clock'' that keeps the story advancing when players do not engage with other authored content.
\item \emph{Repeatable} triggers that can fire indefinitely many times.
\end{itemize}

Second, we plan to implement more flexible trigger detection mechanisms while maintaining system responsiveness. In particular, we may introduce more granular trigger-level settings, such as:
\begin{itemize}
\item An explicitly configurable \emph{cooldown time} between successive activations of a single trigger, allowing authors to ensure that escalating triggers do not ``self-escalate'' by causing characters to say things that almost immediately cause the trigger to fire again.
\item Explicit \emph{trigger ordering constraints} that can be used to gate each trigger's activation on the prior activation (or non-activation) of another trigger. Because it is much quicker to check which triggers have already been activated than to have an LLM evaluate an open-ended classifier rule against the whole story text so far, ordering constraints could allow authors to expand their library of triggers substantially and introduce more intricate potential narrative progressions to a story without overwhelming the system.
\end{itemize}

Finally, we plan to conduct a more comprehensive evaluation study involving having a larger pool of players play through each of the interactive stories created in our initial authoring study. This would enable quantitative analysis of playthrough diversity; responsiveness to player-introduced narrative ideas; trigger accuracy across multiple playthroughs; author satisfaction with different players' experiences of their stories; and other aspects of player experience that we have not yet attempted to gauge here.

\section{Conclusion}

Early-stage testing with six content authors suggests that \textsc{Drama Llama}---a hybrid storylet-based LLM framework for interactive storytelling---can generate coherent and meaningful narratives with believable character interactions. While the system faces challenges around clich\'ed output and LLM stochasticity, we also find that the quality of generated drama closely tracks the level of effort invested in content authoring. This work suggests several promising directions for future research in LLM-based interactive narrative systems, particularly in developing hybrid approaches that enhance authorial control while supporting emergent generation from LLMs. In contrast to the ``dearth of the author'' sometimes experienced around current LLM-supported writing~\cite{Dearth}, and in line with Janet Murray's early writing on interactive narrative, we hope that this work illuminates one path to a future in which ``the hand behind the multiform plot [feels] as firmly present as the hand of the traditional author''~\cite[p. 347]{HamletOnTheHolodeck}---even as improvising machines greatly expand the range of player inputs to which these multiform plots can meaningfully respond.

\ifCLASSOPTIONcaptionsoff
  \newpage
\fi

\bibliographystyle{IEEEtran}
\bibliography{bibliography}

\onecolumn
\appendix 
\subsection{Example Trigger Definition}
\label{sec:exampletriggers}
\noindent
A typical trigger looks like the following:
\begin{verbatim}
{
  "condition": "Has Sepideh noticed Byron withdrawing from the conversation?",
  "actions": [
    "Sepideh raises her voice to ask Byron if he's feeling okay.",
    "Sepideh angrily suggests to Byron that he go upstairs to rest.",
    "Sepideh abruptly grabs Byron's plate and takes it to the kitchen sink."
  ],
  "type": "basic"
}
\end{verbatim}
Each time the drama manager gauges that the \texttt{condition} is met (and that this is the highest-priority trigger whose condition is met on this turn), the next unused entry from \texttt{actions} will be inserted into the story text.

\subsection{System Prompt for Simulation}
\label{sec:simprompt}
\begin{verbatim}
We're writing a story in the form of a play script. The story has these characters:
[Character name 1]: [Description]
[Character name 2]: [Description]
[...]

So far, the script is as follows:
*[world settings]*
[Character name]: [dialogue]
[Character name]: [dialogue]
*[stage action]*
[...]

Suggest a possible next line for the script. Wrap it in <line></line> tags.

\end{verbatim}

\subsection{System Prompt to Check a Single Trigger}
\label{sec:triggerprompt}
\begin{verbatim}
[Same context as Prompt for simulation]

Decide whether the following condition has been met in the script so far:
[Trigger condition]

Return either the single token YES or NO, nothing else.

\end{verbatim}

\subsection{TTCW Results}
\label{ttcw result}
\begin{figure}
    \centering
    \includegraphics[width=1\linewidth]{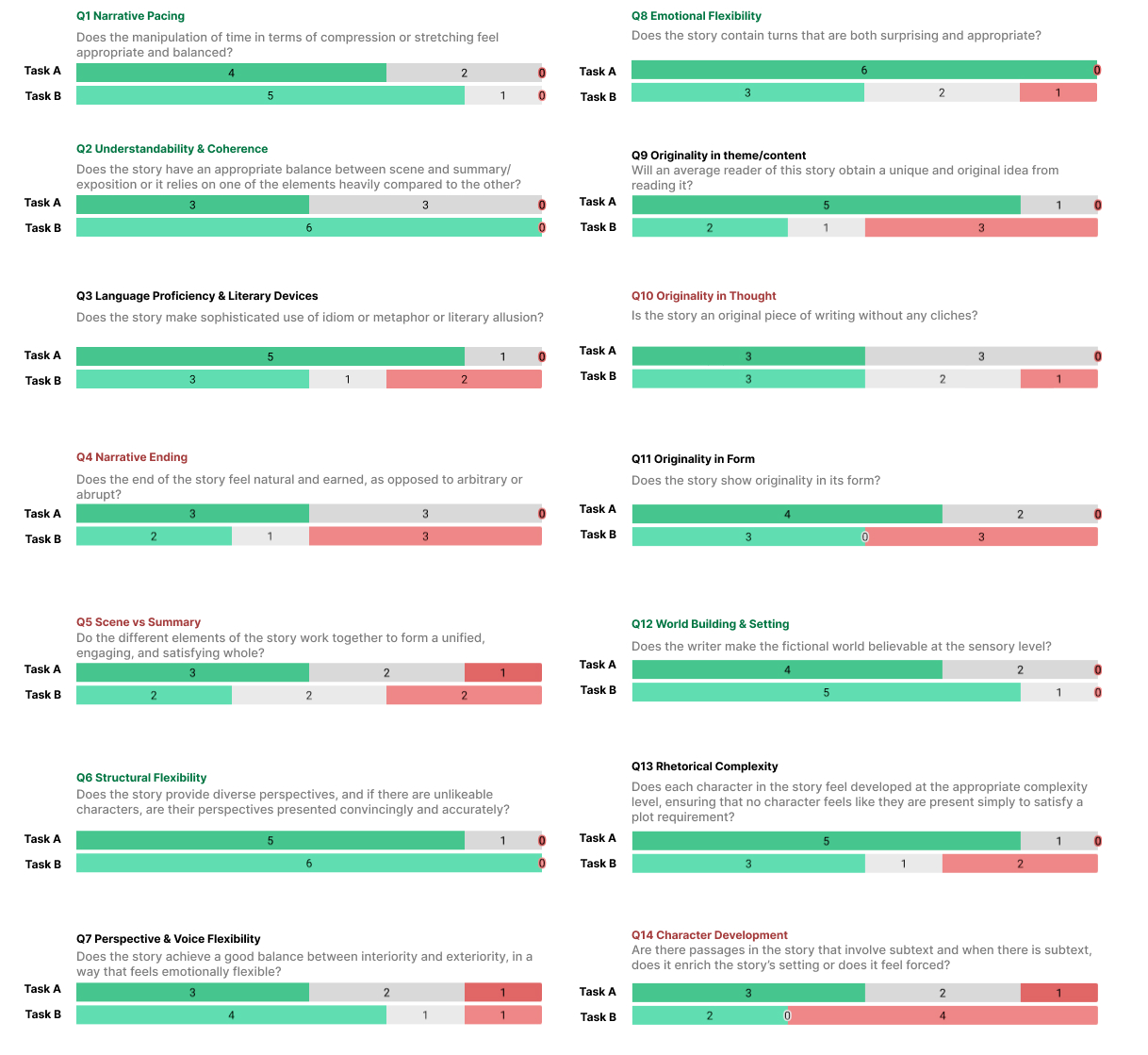}
    \caption{Authors' TTCW evaluations of representative playthroughs of their Task A and Task B stories. For both tasks, authors answered either ``yes'' or ``no'' to each question in the TTCW questionnaire; presented numbers indicate how many authors answered ``yes'' or ``no'' for each question and task.}
\end{figure}

\end{document}